\pdfoutput=1

\documentclass[11pt,twoside,a4paper,cmspaper,final,collab]{cms-tdr}
\def\svnVersion{b71c5be-D}\def\svnDate{2020/03/14}\def\cmsCernNoTag{CERN-EP-2019-288}\def\cmsCernDate{\today}\def\cmsMessage{"Published in Physics Letters B as \href{http://dx.doi.org/10.1016/j.physletb.2020.135345}{\doi{10.1016/j.physletb.2020.135345}.}"}

\begin{document}\cmsNoteHeader{BPH-19-003}

\hyphenation{had-ron-i-za-tion}
\hyphenation{cal-or-i-me-ter}
\hyphenation{de-vices}
\newcommand{\pipi}      {\ensuremath{\PGpp\PGpm}\xspace}
\newcommand{\Lbson}     {\ensuremath{\PGLb\hspace{-.09em}(5912)^0}\xspace}
\newcommand{\Lbstw}     {\ensuremath{\PGLb\hspace{-.09em}(5920)^0}\xspace}
\newcommand{\Lbsth}     {\ensuremath{\PGLb\hspace{-.09em}(6150)^0}\xspace}
\newcommand{\Lbsthl}    {\ensuremath{\PGLb\hspace{-.09em}(6146)^0}\xspace}
\newcommand{\Lbsthh}    {\ensuremath{\PGLb\hspace{-.09em}(6152)^0}\xspace}
\newcommand{\PGSbpm}    {\HepParticle{\PGS}{b}{\pm}\Xspace}
\newcommand{\JPP}       {\ensuremath{\JPsi\pipi}\xspace}
\newcommand{\PJPP}      {\ensuremath{\Pgy\to\JPP}\xspace}
\newcommand{\LbPP}      {\ensuremath{\PGLzb\pipi}\xspace}
\newcommand{\LbPPSS}    {\ensuremath{\PGLzb\PGppm\PGppm}\xspace}
\newcommand{\mLbPP}     {\ensuremath{m_{\PGLzb\pipi}}\xspace}
\newcommand{\mLbPPnos}  {\ensuremath{m_{\PGLzb\PGp\PGp}}\xspace}

\ifthenelse{\boolean{cms@external}}{\providecommand{\cmsLeft}{upper\xspace}}{\providecommand{\cmsLeft}{left\xspace}}
\ifthenelse{\boolean{cms@external}}{\providecommand{\cmsRight}{lower\xspace}}{\providecommand{\cmsRight}{right\xspace}}
\ifthenelse{\boolean{cms@external}}{\providecommand{\figurewidth}{0.98\columnwidth}}{\providecommand{\figurewidth}{0.60\textwidth}}
\ifthenelse{\boolean{cms@external}}{\providecommand{\cmsTable}[1]{#1}}{\providecommand{\cmsTable}[1]{\resizebox{\columnwidth}{!}{#1}}}
\newlength\cmsTabSkip\setlength{\cmsTabSkip}{1ex}

\ifthenelse{\boolean{cms@external}}
{\providecommand{\suppMaterial}{the supplemental material
[URL will be inserted by publisher]}}
{\providecommand{\suppMaterial}{Appendix~\ref{app:suppMat}}}

\cmsNoteHeader{BPH-19-003}
\title{Study of excited  $\PGLzb$ states decaying to $\PGLzb\pipi$ in proton-proton collisions at $\sqrt{s}=13\TeV$}

\date{\today}

\abstract{
A study of excited \PGLzb
baryons is reported,
based on a data sample collected in 2016--2018 with the CMS
detector at the LHC in proton-proton collisions at a center-of-mass energy of 13\TeV, corresponding to an integrated luminosity of up to 140\fbinv.
The existence of four excited \PGLzb states: \Lbson, \Lbstw, \Lbsthl, and \Lbsthh in the $\PGLzb\pipi$ mass spectrum is confirmed,
and their masses are measured.
The $\PGLzb\pipi$ mass distribution exhibits a broad excess of events in the region of 6040--6100\MeV,
whose origin cannot be discerned with the present data.
}

\hypersetup{%
pdfauthor={CMS Collaboration},%
pdftitle={Study of excited Lambda-b statesdecaying to Lambda-b pi+ pi- in pp collisions at sqrt(s)=13 TeV},%
pdfsubject={CMS},%
pdfkeywords={CMS, physics, beauty baryon, heavy baryon, heavy flavor, flavor spectroscopy, experimental results}}

\maketitle

\section{Introduction}
Studies of excited baryonic states are an important aspect of hadron
spectroscopy and help to shed light on the mechanisms responsible for
dynamics of quarks and baryon formation. In particular, spectroscopy of baryons that
contain a heavy-flavor quark, such as the \PGLzb baryon, can test predictions of heavy-quark effective
theory~\cite{HQET}. A number of theoretical calculations exist for various orbital and radial excitations of the ground state
baryons containing a \PQb quark~\cite{Capstick:1986bm,Garcilazo:2007eh,Ebert:2007nw,Roberts:2007ni,Karliner:2008sv,Zhang:2008pm,GarciaRecio:2012db,Yamaguchi:2014era,Torres-Rincon:2014ffa,Chen:2014nyo,Wei:2016jyk,Thakkar:2016dna,Wang:2017goq,Wang:2017kfr,Kawakami:2019hpp}, including those of the \PGLzb baryon. In general, there are a number of excited \PGLzb baryon states
predicted in the 5.9--6.4\GeV mass range. However, predictions are very diverse
in terms of the specific mass spectrum and do not point to any common narrow mass
region in which to search for these excited states. As an additional complication, the widths and the production cross sections of
various excited states are generally unknown. This situation makes experimental searches for excited heavy-quark baryons
both challenging and important for testing various theoretical models.

{\tolerance=800
The existence of two narrow excited $\PGLzb$ states, \Lbson and
\Lbstw, in the $\LbPP$ invariant mass spectrum near the kinematic
threshold was reported by the LHCb Collaboration in 2012~\cite{Aaij:2012da}
(charge-conjugate states are implied throughout this Letter).
The measured masses are $M(\Lbson)=5911.97\pm0.67\MeV$ and $M(\Lbstw)=5919.77\pm0.67\MeV$, and
the respective natural widths were found to be below 0.83 and 0.75\MeV at 95\%
confidence level.
The latter state was confirmed by the CDF Collaboration~\cite{Aaltonen:2013tta} soon thereafter
with the mass measured to be $M(\Lbstw)=5919.22\pm0.76\MeV$.
The precision of these measurements was limited by the large uncertainty in the \PGLzb mass
at the time; the current world-average values $M(\Lbson)=5912.20\pm0.21\MeV$
and $M(\Lbstw)=5919.92\pm0.19\MeV$~\cite{PDG} are based on the updated \PGLzb mass measurement~\cite{Aaij:2014pha,Aaij:2017awb}.
Recently, the LHCb experiment has also presented an observation of
two narrow higher-mass states in the $\LbPP$ spectrum,
with the following masses and widths~\cite{Aaij:2019amv}:
$M(\Lbsthl)=6146.17\pm0.43\MeV$,
$\Gamma(\Lbsthl)=2.9\pm1.3\MeV$, and $M(\Lbsthh)=6152.51\pm0.38\MeV$,
$\Gamma(\Lbsthh)=2.1\pm0.9\MeV$.
\par}

In this Letter, a study of the \LbPP invariant mass distribution in the 5900--6400\GeV range by the CMS
Collaboration is reported.
Both the \Lbson and \Lbstw states near the kinematic threshold are confirmed and their masses are measured.
In addition, the \LbPP distribution is investigated in the higher-mass region
and signals consistent
with the \Lbsthl and \Lbsthh states are observed.
The ground state baryon \PGLzb is reconstructed via its decays into the $\JPsi\Lambda$ and
$\Pgy\Lambda$ channels.
The analysis uses the proton-proton ($\Pp\Pp$) collision data recorded with the CMS detector in 2016--2018, during
the CERN LHC Run 2 at $\sqrt{s} = 13\TeV$, corresponding to an
integrated luminosity of up to 140\fbinv.

\section{The CMS detector}
The central feature of the CMS apparatus is a superconducting solenoid of
6\unit{m} internal diameter, providing a magnetic field of 3.8\unit{T}.
Within the solenoid volume are a silicon pixel and strip tracker, a lead
tungstate crystal electromagnetic calorimeter, and a brass and scintillator
hadron calorimeter, each composed of a barrel and two endcap sections.
Forward calorimeters extend the pseudorapidity coverage provided by the barrel
and endcap detectors. Muons are detected in gas-ionization chambers embedded in
the steel flux-return yoke outside the solenoid. A more detailed description
of the CMS detector, together with a definition of the coordinate system used and
the relevant kinematic variables, can be found in Ref.~\cite{Chatrchyan:2008zzk}.

Events of interest are selected using a two-tiered trigger system~\cite{Khachatryan:2016bia}.
The first level (L1), composed of custom hardware processors, uses information from the
calorimeters and muon detectors to select events at a rate of around 100\unit{kHz} within a time
interval of less than 4\mus. The L1 trigger used in the analysis required at least two muons.
The second level, known as the high-level trigger (HLT), consists of a farm of processors running
a version of the full event reconstruction software optimized for fast processing, and reduces the
event rate to around 1\unit{kHz} before data storage. The set of HLT algorithms used in the analysis
requires two opposite-sign (OS) muons with various pseudorapidity $\eta$ and transverse
momentum \pt thresholds, compatible with being produced in the dimuon decays of \JPsi or \Pgy mesons.
Given that no single trigger algorithm is dedicated to the decay signature of interest, the analysis uses
a combination of several triggers, with integrated luminosities up to 140\fbinv.

\section{Event selection}
The event selection begins by requiring two OS muons passing the
CMS soft-muon identification criteria~\cite{Chatrchyan:2012xi} with
$\pt>3\GeV$ and $\abs{\eta} < 2.2$. The muons must form a common vertex with
a $\chi^2$ fit probability ($P_{\text{vtx}}$) greater than 1\%.
The dimuon invariant mass is required to satisfy $2.90 < M(\MM) < 3.95\GeV$. If $M(\MM)$
is below 3.4\GeV, the dimuon system is considered to be a \JPsi candidate, or a \Pgy candidate otherwise.

Another \Pgy decay channel is also used to increase the signal yield: $\PJPP \to \MM\pipi$.
Two additional, high purity~\cite{Chatrchyan:2014fea}, OS tracks, assumed to be pions and labeled $\PGp_{\Pgy}$,
are required to have $\pt>0.35\GeV$.
They are fit to a common vertex with a \JPsi candidate, using a world-average \JPsi meson mass~\cite{PDG} constraint.
The invariant mass of the \JPsi candidate and the two tracks must satisfy the requirement $3672<M(\JPP)<3700\MeV$,
corresponding to a window centered on the world-average \Pgy meson mass, with a half-width of
approximately three times the corresponding mass resolution.

A \PGL candidate is
formed from a displaced two-prong vertex, assuming the decay $\PGL\to\Pp\PGpm$,
as described in Ref.~\cite{Khachatryan:2010pw}.
The $\Pp\PGpm$ invariant mass is required to be within $\pm 10\MeV$ of the world-average
\PGL baryon mass $m^{\text{PDG}}_{\PGL}$~\cite{PDG}, which corresponds to
approximately three times the \PGL candidate mass resolution. The two tracks are refitted with
their invariant mass constrained to $m^{\text{PDG}}_{\PGL}$, and the obtained
\PGL candidate is required to have $P_{\text{vtx}}>1\%$.

To form the \PGLzb candidates, the \JPsi or \Pgy candidate and the \PGL candidate are fit to a common
vertex with $P_{\text{vtx}}>1\%$,
where the world-average \JPsi or \Pgy mass~\cite{PDG} constraint is applied to the muon pair.
In the case of the $\PJPP$ decay channel, only the $\JPsi \to \MM$ mass constraint is used.

The primary vertex (PV) associated with the \PGLzb candidate is selected among all
the reconstructed vertices by requiring the smallest angle between the
reconstructed \PGLzb candidate momentum and the vector pointing from this vertex to the \PGLzb decay vertex.
The PV is then refitted after removing  the tracks associated with the \PGL and either the \JPsi or \Pgy candidates.
The decay length of the \PGLzb candidate in the transverse plane, $L_{xy}$, is computed as the two-dimensional
distance between the PV and the \PGLzb decay vertex, and is required to exceed
three times its uncertainty. This selection helps to suppress the combinatorial background.
In addition, the transverse momentum of the \PGLzb candidate
is required to be well
aligned with the transverse displacement vector: $\cos\alpha > 0.99$, where $\alpha$ 
is the angle between the projections on the plane transverse to the beams of the \PGLzb 
candidate momentum and of the vector connecting the PV with the \PGLzb decay vertex. 
The numbers of \PGLzb signal
candidates after these requirements are about 39\,000, 3400, and 4300 for the $\JPsi\Lambda$,
$\Pgy\Lambda$ ($\Pgy\to\MM$), and $\Pgy\Lambda$ ($\PJPP$) channels, respectively.

The $\LbPP$ candidates are formed by combining the selected \PGLzb candidates with
two OS tracks originating from the PV, as in
Refs.~\cite{BPH16002, BPH16003, BPH18007}, since the lifetime of excited
\PGLzb states is expected to be negligible, resulting in prompt decays.
Combinations of a \PGLzb candidate with two prompt same-sign (SS) pions are used as
a control channel and form the SS control region, as opposed to the OS signal region.
The higher-\pt pion of the pair is labeled $\PGp^{\pm}_1$ and the lower-\pt pion $\PGp^{\pm}_2$.
To improve the \LbPP invariant mass resolution, 
all tracks forming the PV and the selected \PGLzb candidate, taken as a 
single ``pseudo-track'' with the momentum, its uncertainty, and the assigned mass 
equal to those of the \PGLzb candidate, are refit to a common vertex.
The \LbPP invariant mass $\mLbPP$ is then calculated using the momenta of particles returned by this vertex
fit through the relation $\mLbPP = M(\LbPP) - M(\PGLzb) + m^{\mathrm{PDG}}_{\PGLzb}$,
where $m^{\mathrm{PDG}}_{\PGLzb}=5619.60\pm0.17\MeV$ is the world-average \PGLzb baryon mass~\cite{PDG}.
The PV refitting procedure
improves the \LbPP mass resolution by up to 50\%. Unless specified otherwise,
multiple \LbPP candidates found in the same event are not discarded.

\section{Simulated samples and selection optimization}
Several simulated signal samples with different masses of excited \PGLzb states are used
in the analysis.
The \PYTHIA~8.230 package~\cite{Pythia8} is used to simulate the production of the excited \PGLzb states.
The \PGSzb baryon, with a modified mass value, is used as a proxy
for an excited \PGLzb baryon. The decays are described with \EVTGEN 1.6.0~\cite{Lange2001152}.
Final-state photon radiation is included in \EVTGEN using
\PHOTOS~\cite{BARBERIO1991115,BARBERIO1994291}.
Generated events are then passed to a detailed \GEANTfour-based
simulation~\cite{Agostinelli2003250} of the CMS detector, followed by the
same trigger and reconstruction algorithms used for collision data.
The simulation includes effects from multiple $\Pp\Pp$ interactions in the same
or nearby bunch crossings (pileup) with the multiplicity
distribution matching that observed in data.

Simulated samples are used to optimize the selection criteria using the
Punzi figure of merit~\cite{Punzi:2003bu}, \ie, optimizing the value of
$S/{(\frac{5}{2}+\sqrt{B})}$, where $S$ is the simulated signal yield
and $B$ is the expected background, as estimated using the SS control region.
This optimization scheme is independent of the signal normalization.

The selection requirements are optimized separately for the low-mass
$\mLbPP < 5950\MeV$ and high-mass $5950<\mLbPP<6400\MeV$ regions, using the
 \Lbson and \Lbsth simulated signal samples, respectively.
For the low-mass region, the optimized criteria are:
$\pt(\PGp^{\pm}_2)>0.3\GeV$, $\pt(\pi^{\pm}_1)>0.35\GeV$,
$\cos\alpha>0.995$, $\cos\alpha^{\text{3D}}>0.995$, and $\pt(\PGp_{\Pgy})>0.4\GeV$,
where $\alpha^{\text{3D}}$ is a three-dimensional analog of the angle $\alpha$.
For the high-mass region, the optimized requirements are found to be
$\pt(\Pgp^{\pm}_2)>0.7\GeV$, $\pt(\PGp^{\pm}_1)>1.4\GeV$,
$\pt(\PGLzb)>16\GeV$, $P_{\text{vtx}}(\PGLzb)>2\%$, and $P_{\text{vtx}}(\LbPP)>8\%$.
In the high-mass region, due to higher backgrounds, if multiple
excited \PGLzb candidates in an event pass the above requirements, only the
highest \pt candidate is kept.
In the low-mass region the average number of candidates per event is very close
to one, while in the high-mass region it is around two.

\section{Observed \protect\boldmath\texorpdfstring{$\Lambda^0_{\PQb}\PGp^+\PGp^-$}{Lambda-b pi pi} invariant mass spectra}

\begin{figure}[htb]
    \centering
        \includegraphics[width=\figurewidth]{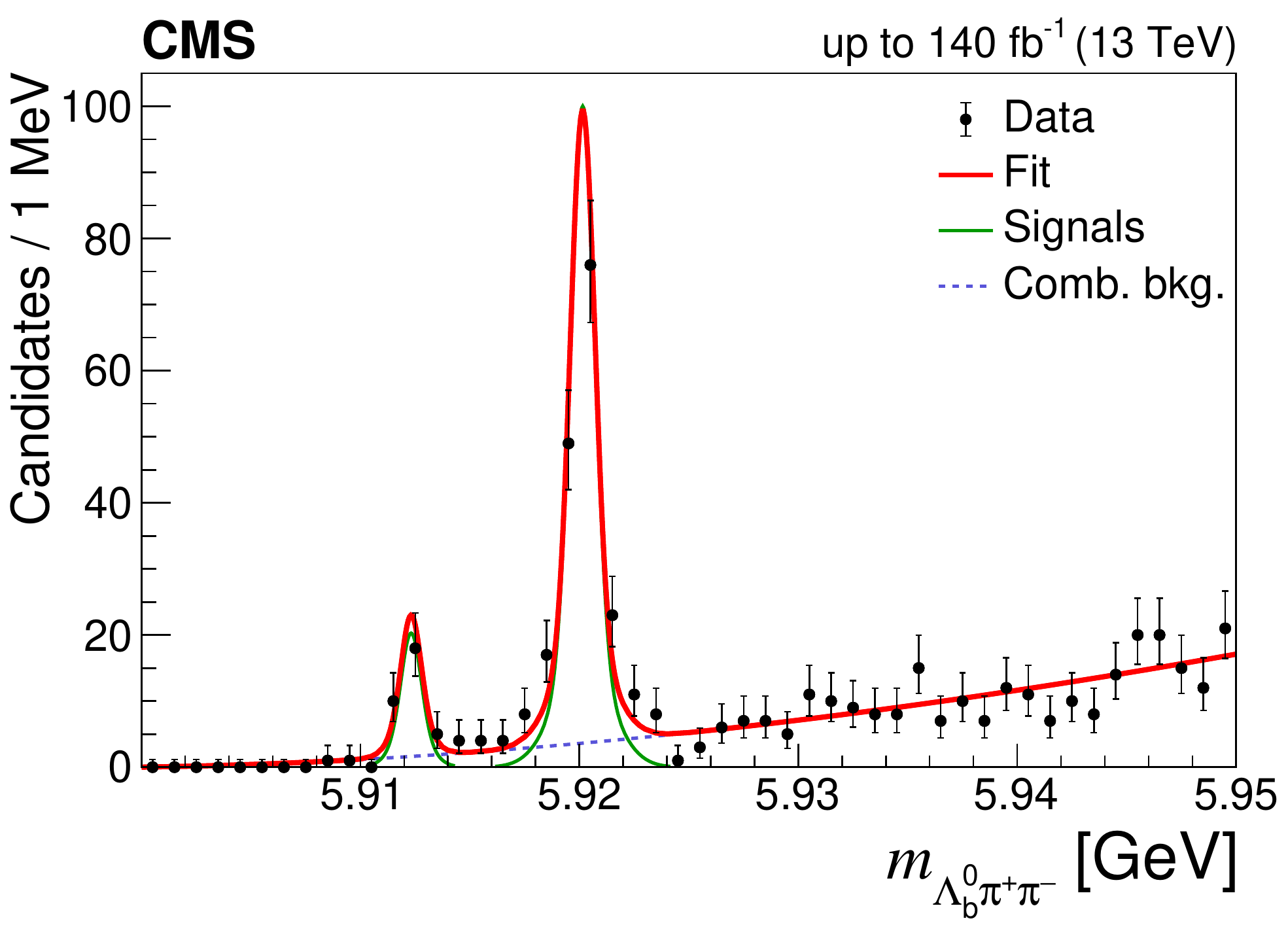}
    \caption{
Invariant mass distribution of the selected \LbPP candidates near threshold.
The vertical bars on the data points display the statistical uncertainties in the data.
The overall fit result is shown by the thick solid line, with the thin and dashed lines
representing the signal and combinatorial background components, respectively.
\label{fig:LbPPlowmass}}
\end{figure}

The observed invariant mass distribution \mLbPP of the selected signal candidates near the threshold is shown in
Fig.~\ref{fig:LbPPlowmass}.
The two narrow peaks corresponding to the \Lbson and \Lbstw baryons are modeled
with double-Gaussian functions with the resolution parameters fixed to those
obtained in simulation (effective resolutions are about 0.6 and 0.8\MeV).
The background is modeled with a threshold
function $(x-x_0)^\beta$, where $x_0$ is the mass threshold value.
The value of $\beta$, as well as the masses and normalizations
of the two signal functions, are free parameters of an unbinned maximum-likelihood fit to data.
The best-fit signal yields are $28.4\pm5.8$ and $159\pm14$ events, and the measured masses
are $5912.32\pm0.12\MeV$ and $5920.16\pm0.07\MeV$, respectively, where the uncertainties
are statistical only. The presence of each of the peaks is established
with a statistical significance of 5.7 and well over 6 standard deviations ($\sigma$),
for the \Lbson and \Lbstw states, respectively, thereby confirming
the existence of these two baryon states. The significances have been evaluated with the
likelihood-ratio technique by applying the one- and two-peak signal hypotheses.
The likelihood ratios are evaluated using an asymptotic formula~\cite{Wilks:1938dza,Cowan:2010js}.
The means and resolution parameters of the two peaks are allowed to vary
in the fit within the Gaussian constraints from Ref.~\cite{PDG} and the simulation.
The significance of the \Lbson state varies between 5.4 and $5.7\sigma$ with the
variations in the fit model used to estimate the systematic uncertainties,
as detailed in Section~\ref{sec:syst}; the significance of the \Lbstw state
remains well above $6\sigma$.

\begin{figure}[htb]
    \centering
        \includegraphics[width=\figurewidth]{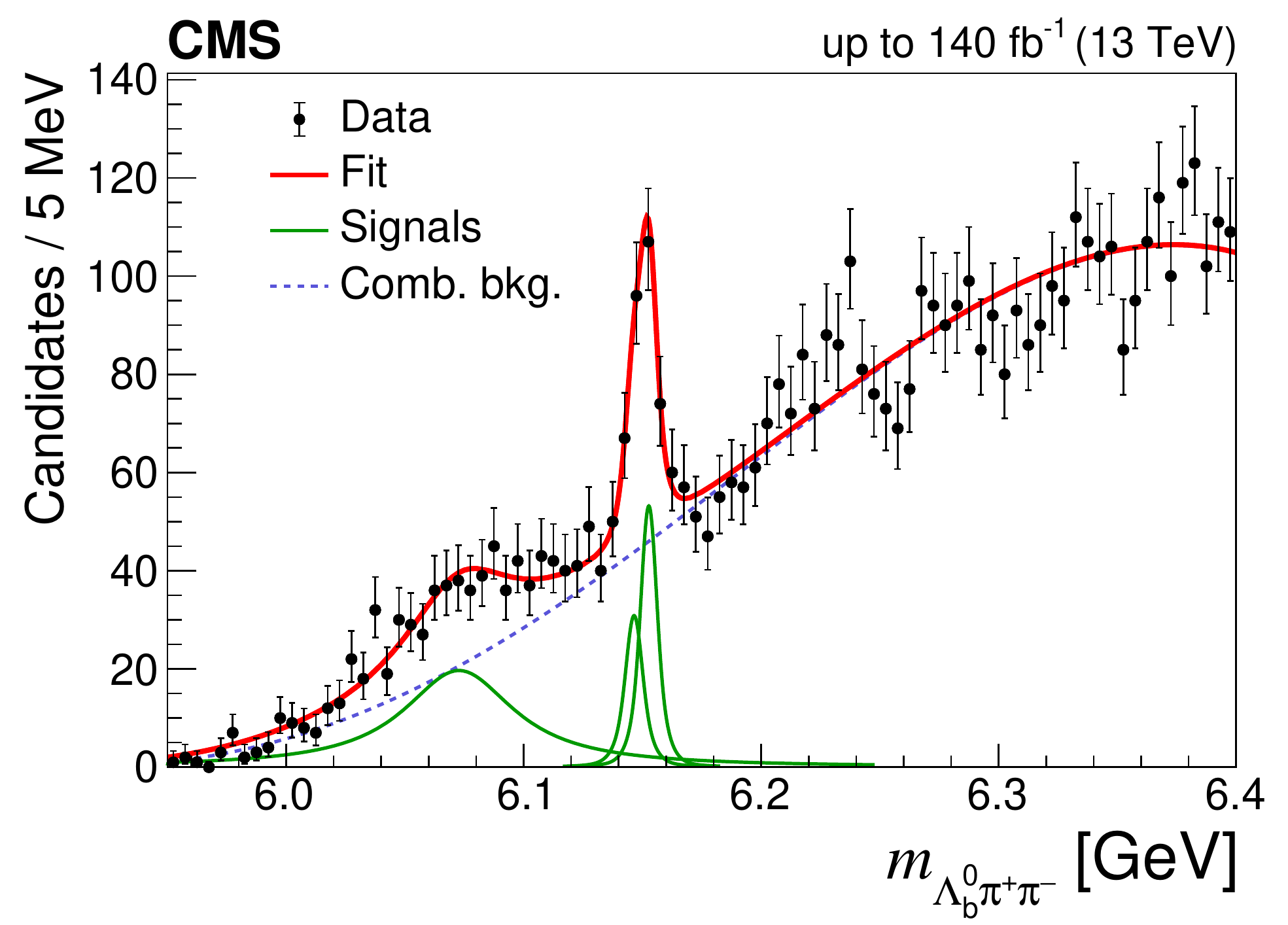}
    \caption{
Invariant mass distribution of the selected \LbPP candidates in the high-mass region.
The vertical bars on the data points represent the statistical uncertainties in the data.
The overall fit result is shown by the thick solid line. The thin lines present the
contributions from the two signal peaks and the broad enhancement. The dashed line displays
the combinatorial background.
\label{fig:LbPPhighmass}}
\end{figure}

Higher masses in the \mLbPP distribution are studied as well, as shown in
Fig.~\ref{fig:LbPPhighmass}.  A narrow peak at approximately 6150\MeV is evident, consistent with an
overlap of the \Lbsthl and \Lbsthh signals, as well as a broad enhancement
in the region below 6100\MeV.
None of these features are present in the SS control region, as shown
in \suppMaterial.

A number of cross-checks have been performed to understand if the broad enhancement
can be the result of a kinematic reflection or produced by a background process. It was found that the enhancement
is not compatible with the partially reconstructed decays of \Lbsthl or \Lbsthh
states into $\PGLzb\pipi\PGpz$ (where the \PGpz is lost). To check if it
can be due to some other state decaying into the $\PGLzb\PKpm\PGpmp$ channels,
the $\PGLzb\PK\PGp$ invariant mass distributions are obtained by substituting the pion
mass with the kaon mass. No significant enhancements over the smooth
background are found. The \mLbPP background distribution is found to be
in agreement between the SS and OS regions in the simulation and does not
show any enhancement in the 6000--6100\MeV mass region.
The two-dimensional distributions of the \LbPP mass versus the $\PGLzb\PGpp$
and $\PGLzb\PGpm$ masses from data are shown in \suppMaterial.
If the $\PGLzb\PGppm$ invariant mass ranges corresponding to the \PGSmb, \PGSpb, \PGSstmb, and \PGSstpb baryons
are vetoed, the SS and OS mass distributions in data are found to be in
agreement in the region below 6100\MeV and do not exhibit a broad enhancement,
as shown in \suppMaterial. This suggests that the broad excess might be
related to the intermediate \PGSbpm and $\PGSstb^\pm$ baryon states,
although the current size of the data set does not allow this hypothesis to be tested.

The observed \mLbPP distribution in the high-mass region
is fit with a sum of three signal functions and
a smooth background function obtained by multiplying the threshold function $(x-x_0)^\beta$ by a first-order polynomial.
The signal function describing the broad structure below 6100\MeV is
a single Breit--Wigner function convolved with a double-Gaussian resolution
function obtained from simulation.
The narrow peak around 6150\MeV is modeled with the sum of two Breit--Wigner
functions, each convolved with a double-Gaussian resolution function obtained from simulation,
having an effective mass resolution of about 3.8\MeV. The natural widths of the two signals
are fixed to those measured by the LHCb Collaboration~\cite{Aaij:2019amv}.
The fit results for the yields and masses, respectively, are
$301\pm72$ and $6073\pm5\MeV$ for the broad enhancement,
$70\pm35$ and $6146.5\pm1.9\MeV$ for the \Lbsthl, and
$113\pm35$ and $6152.7\pm1.1\MeV$ for the \Lbsthh.
The measured natural width of the broad excess is $55\pm11\stat\MeV$. 
While this work was under the journal review, a similar structure, consistent 
with the one reported here and possibly a new baryon state, has been observed by the LHCb 
Collaboration~\cite{Aaij:2020rkw}.

Using the likelihood-ratio technique and the one- versus two-peak hypotheses,
the presence of two peaks has a statistical significance of $0.4\sigma$,
indicating that the data are also consistent with a single peak at 6150\MeV.
For the double-peak hypothesis, the natural widths of the two states are allowed to vary in the fit
within the Gaussian constraints from the LHCb measurement~\cite{Aaij:2019amv}.
In the single-peak hypothesis, the mass and the natural width of the signal peak are free parameters of the fit.
In both cases, the mass resolution is allowed to float in the fit within its Gaussian uncertainty estimated from simulation.
The local statistical significance of the single-peak hypothesis with respect to the background-only hypothesis is
found to be over $6\sigma$ in the baseline fit, and varies between 5.4 and $6.5\sigma$
with the changes in the fit range and the model used to estimate the systematic
uncertainties, as detailed in Section~\ref{sec:syst}.
The broad enhancement has a local statistical significance of about $4\sigma$.
Resonances with masses between 6200 and 6400\MeV have been also considered in the fit model
and no significant excess was found.
The present amount of data does not allow us to perform conclusive studies of 
$\PGSstb^\pm\to\PGLzb\PGppm$ decay contributions to the resonances decaying 
into \LbPP.

\section{Systematic uncertainties\label{sec:syst}}
Several sources of systematic uncertainties in the measured masses are
considered. To evaluate the systematic uncertainties related to the choice of the fit model,
several alternative signal and background functions are tested.
Uncertainties related to the choice of the signal and background models are
evaluated separately.
The systematic uncertainty in each measurement is calculated as the
maximum deviation of the observed mass value from the baseline
fit result.
The alternative signal model corresponds to a single-Gaussian resolution function;
the alternative background models for the low- and high-mass regions are first- and second-order
polynomials, respectively, multiplied by the same threshold function as in the baseline fit.

For the high-mass region, the nature of the broad excess below 6100\MeV
is unclear, therefore an additional fit is performed in the
region $\mLbPP>6100\MeV$, and the observed deviations from the baseline
fit result in the measured masses are taken as
the systematic uncertainties related to the possible presence of the broad resonance.

The systematic uncertainty from the choice of the fit range is evaluated by
extending the range up to 6650\MeV. The observed deviations in the measured masses
are taken as the systematic uncertainties.
The systematic uncertainties due to fit range variations are negligible
for the \Lbson and \Lbstw states.

In the baseline fits of the \mLbPP distributions, the mass resolutions are fixed
to those estimated from simulated event samples. The systematic uncertainty associated
with a possible
differences between data and simulation is calculated using the following procedure.
The mass resolutions are compared between data and simulation for the
copious $\PGLzb\to\JPsi\PGL$ signal: they are, respectively, 15.25
and 15.78\MeV, corresponding to a difference of 3.5\%. This difference is considered to be
the uncertainty in the resolution due to the data-simulation difference.
To estimate the effect of this uncertainty on the measured masses, the baseline fits are redone
with the resolutions increased or decreased by 3.5\%, and the largest deviation
in the measured masses from the baseline fit results is considered as the
systematic uncertainty due to the mass resolution.

The measured masses of the \Lbsthl and \Lbsthh states have an additional systematic
uncertainty due to the fact that their natural widths were fixed in the nominal fit to the values
reported by LHCb. To estimate the respective uncertainty, the nominal fit
is repeated with the natural widths fixed to the central values obtained by LHCb plus or minus the
corresponding uncertainties (in total 8 additional fits are performed).

A potential bias in the mass measurement due to a possible misalignment of the
tracker detectors has been evaluated by comparing
distributions obtained in 2016, 2017, and 2018 running periods, which is a reasonable
comparison, given that an important fraction of the CMS tracking detector was replaced
between the 2016 and 2017 data taking. As expected, the alignment of the detector leads to
a negligible systematic uncertainty in the results reported in this Letter.

The various systematic uncertainties are summarized in
Table~\ref{tab:syst}, together with the total uncertainties calculated
as the quadratic sum of the individual sources.

\begin{table*}[t]
\centering
\topcaption{
Systematic uncertainties (in \MeV) in the measured masses. A dash means that the
corresponding uncertainty is negligible, and ``N/A" means that it does not apply.
\label{tab:syst}}
\cmsTable{
\begin{tabular}{lcccc}
        Source
    &   $M(\Lbson)$
    &   $M(\Lbstw)$
    &   $M(\Lbsthl)$
    &   $M(\Lbsthh)$
\\
\hline
        Signal model
    &   $0.005$
    &   $0.011$
    &   $0.21$
    &   $0.23$
\\
        Background model
    &   $0.004$
    &   ---
    &   $0.16$
    &   $0.14$
\\
        Inclusion of the broad excess region
    &   N/A
    &   N/A
    &   $0.35$
    &   $0.14$
\\
        Fit range
    &   ---
    &   ---
    &   $0.40$
    &   $0.02$
\\
        Mass resolution
    &   $0.007$
    &   $0.001$
    &   $0.01$
    &   $0.09$
\\
        Knowledge of $\Gamma$
    &   N/A
    &   N/A
    &   $0.43$
    &   $0.26$
\\[\cmsTabSkip]
        Total
    &   $0.009$
    &   $0.011$
    &   $0.77$
    &   $0.41$
\end{tabular}
}
\end{table*}

\section{Summary}
In summary, using the $\Pp\Pp$ collision data recorded with the CMS detector at
$\sqrt{s} = 13\TeV$, corresponding to an integrated luminosity of up to 140\fbinv,
the existence of the \Lbson and \Lbstw baryons is confirmed. Their masses, with
respect to the \PGLzb mass, are measured to be
$292.72\pm0.12\pm0.01\MeV$
and
$300.56\pm0.07\pm0.01\MeV$,
respectively, where the first uncertainty is statistical
and the second is systematic. By adding the known \PGLzb mass of
$5619.60\pm0.17\MeV$~\cite{PDG}, we report the mass measurements
\begin{linenomath}\begin{equation*}\begin{aligned}
M(\Lbson)&=5912.32\pm0.12\pm0.01\pm0.17\MeV,\\
M(\Lbstw)&=5920.16\pm0.07\pm0.01\pm0.17\MeV,
\end{aligned}\end{equation*}\end{linenomath}
where the third uncertainty is the uncertainty in the world-average \PGLzb mass.
The obtained values are consistent with the world-average values and have similar precision.

In addition, the \LbPP invariant mass spectrum is investigated in the mass
range up to 6400\MeV.
A narrow peak is observed with a mass close to 6150\MeV, with a significance
over 5 standard deviations, consistent with the superposition of the
\Lbsthl and \Lbsthh baryons recently observed by the LHCb Collaboration~\cite{Aaij:2019amv}.
The masses of these states are measured to be
\begin{linenomath}\begin{equation*}\begin{aligned}
M(\Lbsthl) &= 6146.5\pm1.9\pm 0.8\pm0.2\MeV,\\
M(\Lbsthh) &= 6152.7\pm1.1\pm 0.4\pm0.2\MeV,
\end{aligned}\end{equation*}\end{linenomath}
where the first uncertainty is statistical, the second is systematic, and
the third is the uncertainty in the world-average \PGLzb mass value. The
corresponding mass differences with respect to the \PGLzb mass are
\begin{linenomath}\begin{equation*}\begin{aligned}
M(\Lbsthl)-M(\PGLzb) &= 526.9\pm1.9\pm0.8\MeV,\\
M(\Lbsthh)-M(\PGLzb) &= 533.1\pm1.1\pm0.4\MeV.
\end{aligned}\end{equation*}\end{linenomath}
These measurements are not as precise as, but are in good agreement with the LHCb results~\cite{Aaij:2019amv}.

In addition, a broad excess of events is observed in the region 6040--6100\MeV, not
present in the same-sign \LbPPSS distribution.
If it is fit with a single Breit-Wigner function, the returned mass and width are
$6073\pm5\stat\MeV$ and $55\pm11\stat\MeV$. 
However, it is not excluded that this enhancement is an overlap of more
than one state with close masses or is created by
the partially reconstructed decays of higher-mass states.
More data are needed to elucidate the nature of this excess.

\begin{acknowledgments}
We congratulate our colleagues in the CERN accelerator departments for the excellent performance of the LHC and thank the technical and administrative staffs at CERN and at other CMS institutes for their contributions to the success of the CMS effort. In addition, we gratefully acknowledge the computing centers and personnel of the Worldwide LHC Computing Grid for delivering so effectively the computing infrastructure essential to our analyses. Finally, we acknowledge the enduring support for the construction and operation of the LHC and the CMS detector provided by the following funding agencies: BMBWF and FWF (Austria); FNRS and FWO (Belgium); CNPq, CAPES, FAPERJ, FAPERGS, and FAPESP (Brazil); MES (Bulgaria); CERN; CAS, MoST, and NSFC (China); COLCIENCIAS (Colombia); MSES and CSF (Croatia); RPF (Cyprus); SENESCYT (Ecuador); MoER, ERC IUT, PUT and ERDF (Estonia); Academy of Finland, MEC, and HIP (Finland); CEA and CNRS/IN2P3 (France); BMBF, DFG, and HGF (Germany); GSRT (Greece); NKFIA (Hungary); DAE and DST (India); IPM (Iran); SFI (Ireland); INFN (Italy); MSIP and NRF (Republic of Korea); MES (Latvia); LAS (Lithuania); MOE and UM (Malaysia); BUAP, CINVESTAV, CONACYT, LNS, SEP, and UASLP-FAI (Mexico); MOS (Montenegro); MBIE (New Zealand); PAEC (Pakistan); MSHE and NSC (Poland); FCT (Portugal); JINR (Dubna); MON, RosAtom, RAS, RFBR, and NRC KI (Russia); MESTD (Serbia); SEIDI, CPAN, PCTI, and FEDER (Spain); MOSTR (Sri Lanka); Swiss Funding Agencies (Switzerland); MST (Taipei); ThEPCenter, IPST, STAR, and NSTDA (Thailand); TUBITAK and TAEK (Turkey); NASU (Ukraine); STFC (United Kingdom); DOE and NSF (USA).

\hyphenation{Rachada-pisek} Individuals have received support from the Marie-Curie program and the European Research Council and Horizon 2020 Grant, contract Nos.\ 675440, 752730, and 765710 (European Union); the Leventis Foundation; the A.P.\ Sloan Foundation; the Alexander von Humboldt Foundation; the Belgian Federal Science Policy Office; the Fonds pour la Formation \`a la Recherche dans l'Industrie et dans l'Agriculture (FRIA-Belgium); the Agentschap voor Innovatie door Wetenschap en Technologie (IWT-Belgium); the F.R.S.-FNRS and FWO (Belgium) under the ``Excellence of Science -- EOS" -- be.h project n.\ 30820817; the Beijing Municipal Science \& Technology Commission, No. Z191100007219010; the Ministry of Education, Youth and Sports (MEYS) of the Czech Republic; the Deutsche Forschungsgemeinschaft (DFG) under Germany’s Excellence Strategy -- EXC 2121 ``Quantum Universe" -- 390833306; the Lend\"ulet (``Momentum") Program and the J\'anos Bolyai Research Scholarship of the Hungarian Academy of Sciences, the New National Excellence Program \'UNKP, the NKFIA research grants 123842, 123959, 124845, 124850, 125105, 128713, 128786, and 129058 (Hungary); the Council of Science and Industrial Research, India; the HOMING PLUS program of the Foundation for Polish Science, cofinanced from European Union, Regional Development Fund, the Mobility Plus program of the Ministry of Science and Higher Education, the National Science Center (Poland), contracts Harmonia 2014/14/M/ST2/00428, Opus 2014/13/B/ST2/02543, 2014/15/B/ST2/03998, and 2015/19/B/ST2/02861, Sonata-bis 2012/07/E/ST2/01406; the National Priorities Research Program by Qatar National Research Fund; the Ministry of Science and Education, grant no. 14.W03.31.0026 (Russia); the Programa Estatal de Fomento de la Investigaci{\'o}n Cient{\'i}fica y T{\'e}cnica de Excelencia Mar\'{\i}a de Maeztu, grant MDM-2015-0509 and the Programa Severo Ochoa del Principado de Asturias; the Thalis and Aristeia programs cofinanced by EU-ESF and the Greek NSRF; the Rachadapisek Sompot Fund for Postdoctoral Fellowship, Chulalongkorn University and the Chulalongkorn Academic into Its 2nd Century Project Advancement Project (Thailand); the Kavli Foundation; the Nvidia Corporation; the SuperMicro Corporation; the Welch Foundation, contract C-1845; and the Weston Havens Foundation (USA). \end{acknowledgments}

\bibliography{auto_generated}
\ifthenelse{\boolean{cms@external}}{}{
\clearpage
\appendix
\numberwithin{table}{section}
\numberwithin{figure}{section}
\section{Additional studies of the broad enhancement\label{app:suppMat}}
The measured $m_{\PGLzb\PGp\PGp}$ distribution is compared between the OS signal
and SS control regions in Fig.~\ref{fig:FigApp1}.
\begin{figure}[htb]
    \centering
        \includegraphics[width=\figurewidth]{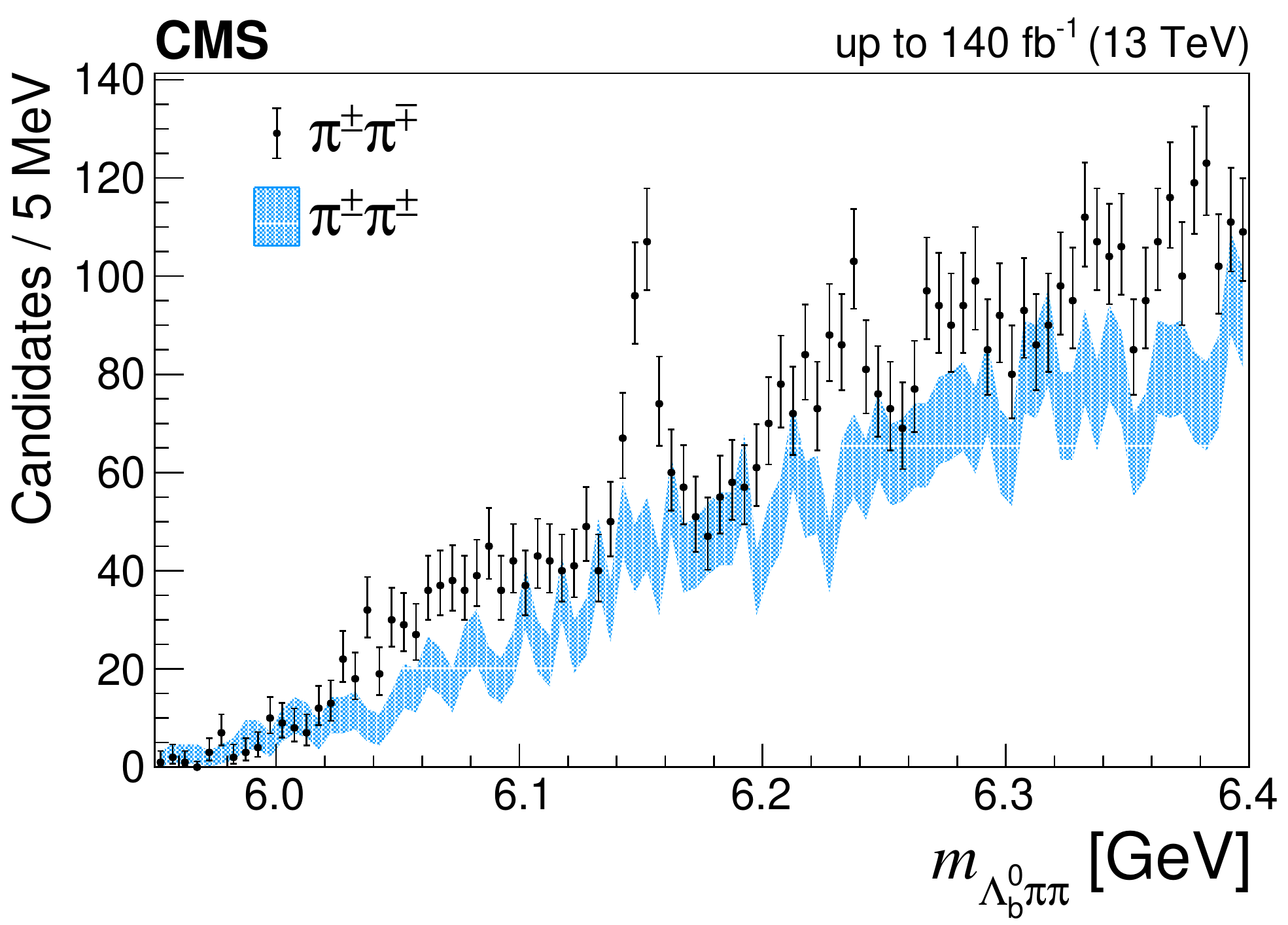}
    \caption{
Comparison of the invariant mass distribution of the selected $\PGLzb\PGp\PGp$
candidates in the SS background (band) and OS signal (points) channels.
The vertical bars on the OS data points and the width of the bands for
the SS distribution indicate the statistical uncertainties only.
\label{fig:FigApp1}}
\end{figure}

The measured two-dimensional distributions of the $\PGLzb\PGpp\PGpm$ mass versus
the $\PGLzb\PGpp$ (\cmsLeft) and $\PGLzb\PGpm$ (\cmsRight) masses,
in the range $\mLbPP<6100\MeV$, are shown in Fig.~\ref{fig:FigApp2}.
\begin{figure}[htb]
    \centering
        \includegraphics[width=0.48\textwidth]{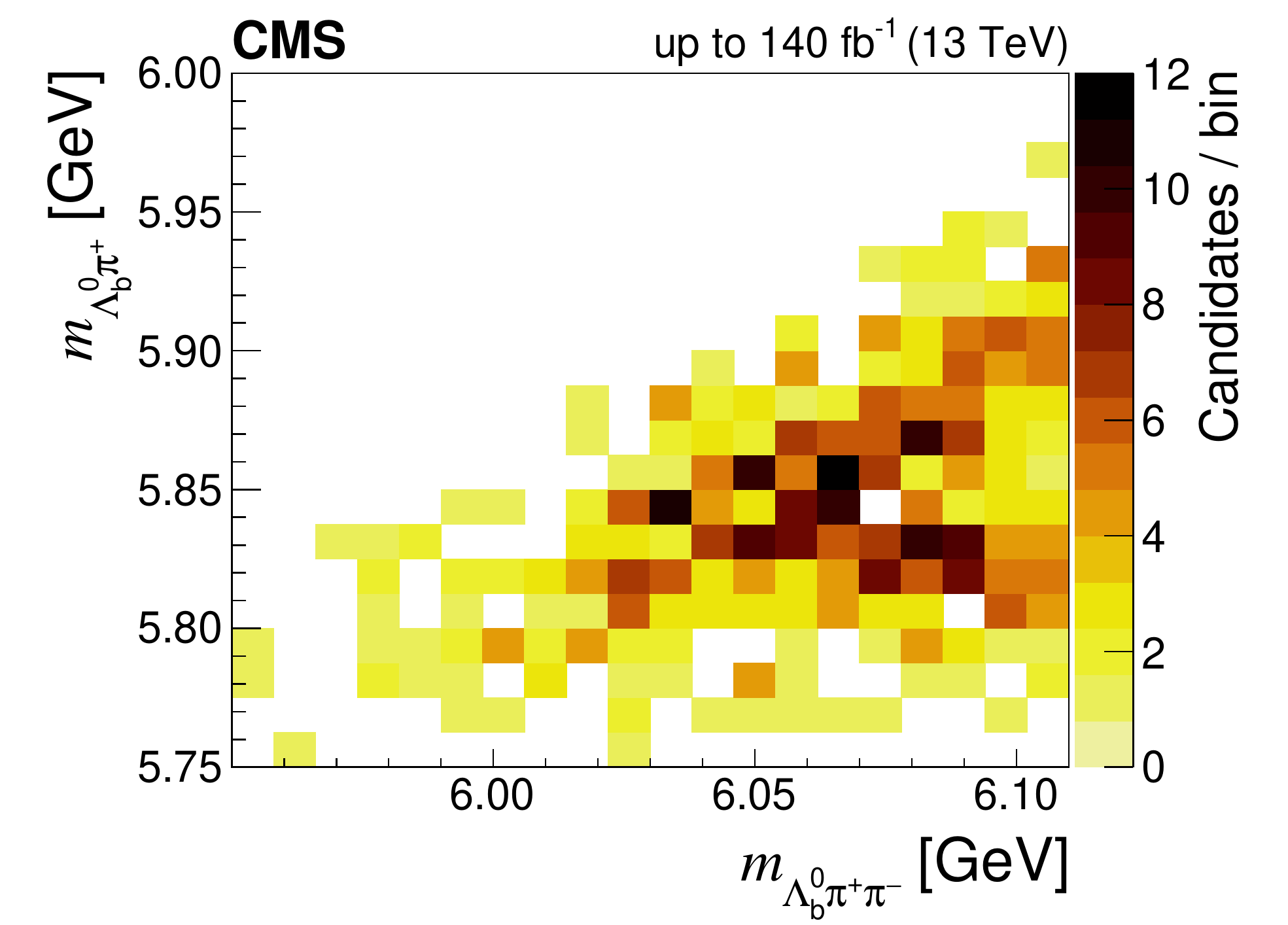}
        \includegraphics[width=0.48\textwidth]{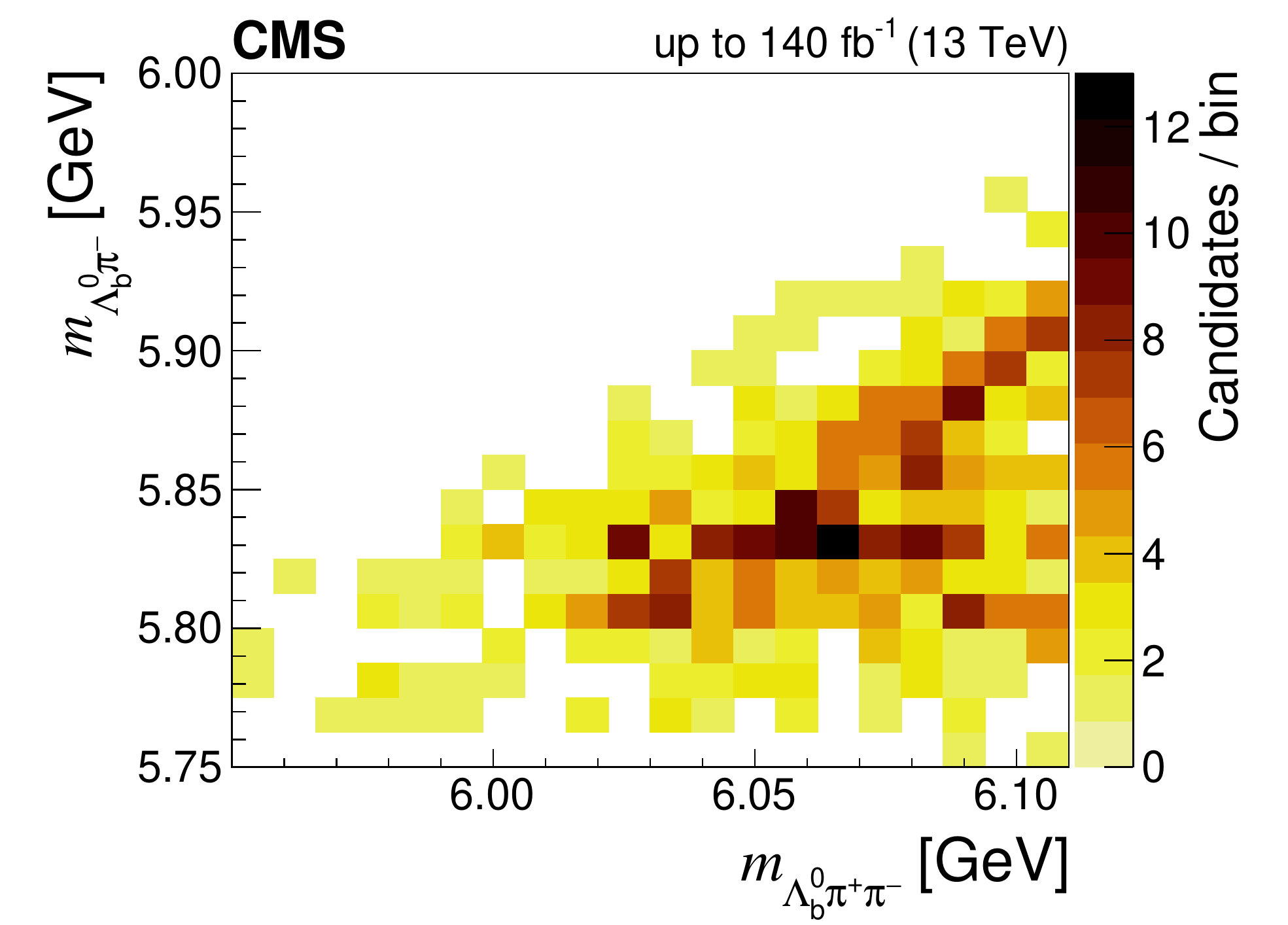}
    \caption{ Two-dimensional distribution of the $\PGLzb\PGpp\PGpm$ mass versus the
     $\PGLzb\PGpp$ mass (\cmsLeft) and  $\PGLzb\PGpm$ mass (\cmsRight), in the range $\mLbPP<6100\MeV$.
    The scale on the right of each plot gives the number of candidates per $8\MeV\times12.5\MeV$ bin.
    \label{fig:FigApp2}}
\end{figure}

Figure~\ref{fig:FigApp3} shows the observed \mLbPPnos distribution
for OS (data points) and SS (band) candidates after possible
contributions from the \PGSbpm and $\PGSstb^\pm$ baryon decays into
$\PGLzb\PGppm$ are vetoed: in addition to the baseline selection requirements,
a $\PGLzb\PGp\PGp$ candidate is discarded if the $\PGLzb\PGpp$ mass falls
in the $[5800,\,5845]\MeV$ region, or the $\PGLzb\PGpm$ mass falls in the
$[5800,\,5850]\MeV$ region.
\begin{figure}[htb]
    \centering
        \includegraphics[width=\figurewidth]{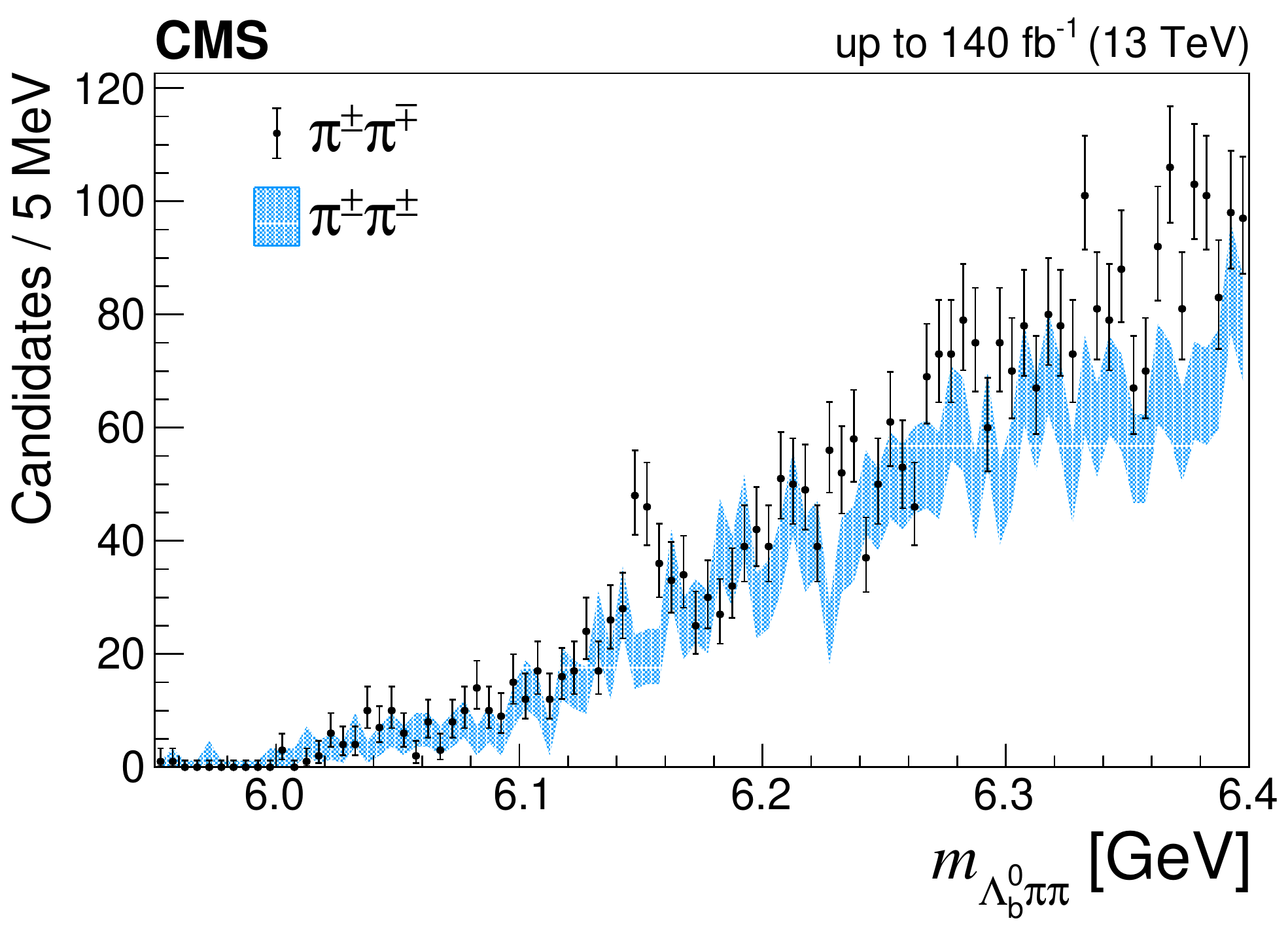}
    \caption{The distribution of \mLbPPnos for OS (points) and SS (band) candidates after
    the possible \PGSbpm and $\PGSstb^\pm$ baryon contributions in the $\PGLzb\PGppm$
    mass spectrum are vetoed. The vertical bars on the OS data points and the width of the
    band for the SS distribution indicate the statistical uncertainties only.
    \label{fig:FigApp3}}
\end{figure}

}

\cleardoublepage \section{The CMS Collaboration \label{app:collab}}\begin{sloppypar}\hyphenpenalty=5000\widowpenalty=500\clubpenalty=5000\vskip\cmsinstskip
\textbf{Yerevan Physics Institute, Yerevan, Armenia}\\*[0pt]
A.M.~Sirunyan$^{\textrm{\dag}}$, A.~Tumasyan
\vskip\cmsinstskip
\textbf{Institut f\"{u}r Hochenergiephysik, Wien, Austria}\\*[0pt]
W.~Adam, F.~Ambrogi, T.~Bergauer, M.~Dragicevic, J.~Er\"{o}, A.~Escalante~Del~Valle, M.~Flechl, R.~Fr\"{u}hwirth\cmsAuthorMark{1}, M.~Jeitler\cmsAuthorMark{1}, N.~Krammer, I.~Kr\"{a}tschmer, D.~Liko, T.~Madlener, I.~Mikulec, N.~Rad, J.~Schieck\cmsAuthorMark{1}, R.~Sch\"{o}fbeck, M.~Spanring, W.~Waltenberger, C.-E.~Wulz\cmsAuthorMark{1}, M.~Zarucki
% ...
\vskip\cmsinstskip
\dag: Deceased\\
1:  Also at Vienna University of Technology, Vienna, Austria\\
% ...
\end{sloppypar}
\end{document}